\documentclass[useAMS]{mn2e}
\input epsf

\def\BE{\begin{equation}}
\def\EE{\end{equation}}

\def\BA{\begin{eqnarray}}
\def\EA{\end{eqnarray}}

\title[Modelling the components of binaries in Hyades]{Modelling the components of binaries in Hyades: The dependence of the mixing-length parameter on stellar mass}
\author[M. Y{\i}ld{\i}z et al.]{M. Y{\i}ld{\i}z$^{1}$\thanks{E-mail:
mutlu.yildiz@ege.edu.tr},
K.Yakut$^{1,2}$,
H. Bak\i\c{s}$^{3}$ and
A. Noels$^{4}$  \\
$^{1}$Ege University, Department of Astronomy and Space Sciences, Bornova, 35100 \.Izmir, Turkey\\
$^{2}$Institute of Astronomy, Catholic University of Leuven, Celestijnenlaan 200 B, 3001 Leuven, Belgium\\
$^{3}$Department of Physics, Faculty of Arts and Sciences, \c{C}anakkale Onsekiz Mart University, 17100 \c{C}anakkale, Turkey\\
$^{4}$Institut d'Astrophysique et G\'{e}ophysique, Universit\'{e} de Li\`{e}ge, All\'{e}e du 6 A\^{out}, B-4000 Li\`{e}ge, Belgium}

\begin{document}

\date{.....}

\pagerange{\pageref{firstpage}--\pageref{lastpage}} \pubyear{2005}

\maketitle
\label{firstpage}

\begin{abstract}

We present our findings based on a detailed analysis for the binaries of the Hyades,
in which the masses of the components are well known.
We fit the models of components of a binary system to the observations so as to give the observed
total $V$ and $B-V$ of that system and the observed slope of the main-sequence in the corresponding parts.
According to our findings, there is a very definite
relationship between the mixing-length parameter and the stellar mass. The fitting formula for this
relationship can be given  as $\alpha = 9.19 (M/M_\odot-0.74)^{0.053}-6.65$, which is valid for
stellar masses greater than $0.77M_\odot$.
While no strict information is gathered for the chemical composition of the cluster, as a result
of degeneracy in the colour-magnitude diagram, by adopting $Z=0.033$ and using models
for the components of 70 Tau and $\theta^2$ Tau we find the hydrogen abundance to be $X=0.676$ and the age to be 670 Myr.
If we assume that $Z=0.024$, then $X=0.718$ and the age is 720 Myr. Our findings concerning the
mixing length parameter are valid for both sets of the solution. { For both components of the active binary system
V818 Tau, the differences between radii of the models with $Z=0.024$ and the observed radii are only about 4 percent.
More generally, the effective temperatures of the models of low mass stars in the binary systems
studied are in good agreement with those determined by spectroscopic methods.}

\end{abstract}
\begin{keywords}
stars: interior -- stars: evolution -- stars: individual: V818 Tau; HD 27149; 70 Tau; 51 Tau;
$\theta^2$ Tau
-- stars: abundances -- binaries: eclipsing -- open clusters and associations: individual: Hyades
\end{keywords}

\section{Introduction}
Open clusters and binary systems have their own essential roles in almost all branches of astrophysics.
While we obtain information on the fundamental properties (such as mass,
radius and luminosity) of components of binary systems from their observation (see e.g., Andersen 1991),
the age of stars of an open cluster can be taken
fairly accurately as the MS life time of its brightest
($normal$) MS star. The age and the fundamental properties of a system are complementary to each other
for the purposes of a better understanding of stellar structure and evolution. Therefore, binary systems in
clusters are invaluable. In this respect, the Hyades
open cluster is an absolute treasure: in addition to very precise observational data for the distance
and photometric measurements of its members (de Bruijne et al. 2001 and Perryman et al. 1998), the masses of the components of its five double-lined binaries are also known. These binaries are
V818 Tau (Peterson \& Solensky, 1988), 70 Tau (Fin 342), 51 Tau, $\theta^2$ Tau and $\theta^1$ Tau (Torres et al. 1997a, 1997b, 1997c).

The cluster itself and its binaries have been the subject of innumerable papers. Recently,
Pinsonneault et al. (2003), Lebreton et al. (2001) and
Lastennet et al. (1999) have researched (some or all of)
these binaries in detail by comparing the observational results with models
for the internal structure
of the component stars. Lastennet et al. (1999) tested the validity of three independent sets of stellar
evolutionary tracks, using good photometric data of V818 Tau, 51 Tau and $\theta^2$ Tau.
Lebreton et al. (2001) focused on the determination of the helium abundance by
considering in detail all of { the five binaries}, and derived the helium abundance as $Y=0.255$,
while Pinsonneault et al. (2003) found $Y=0.271$ from the calibrating the components of the eclipsing binary
V818 Tau. Both of the studies stress the difficulty of calibration of the radii of the components of V818 Tau.
Furthermore, Lebreton et al. (2001) take note of the stellar mass dependence of the mixing-length parameter
($\alpha$).

{
The mixing-length parameter $\alpha=l/H_P$ is an unknown in stellar modelling and is often chosen to be constant
and equal to the solar value.
As discussed in many papers, however,
there is no good reason for keeping it constant; it may change from star to star (see below) and form phase to phase
(Castellani et al. 2001; Chieffi et al. 1995).

For the stellar mass dependence of $\alpha$, contradictory results are obtained from studies on
binaries by different investigators. In studies on eclipsing binaries (Lastennet et al. 2003;
Ludwig and Salaris 1999; Lebreton et al. 2001),
it is reported that $\alpha$ is an increasing function of the stellar mass.
Such a dependence is also found by Morel et al. (2000) in their studies on the visual binary $\iota$ Peg.
However, the other two possibilities are also reported for the visual binaries: while Fernandes et al. (1998) state that
the solar value can be used to model components of four visual binaries they studied, Pourbaix et al. (1999)
find from the calibration of $\alpha$ Cen that $\alpha$ is a decreasing function of the stellar mass.
This complexity for the visual binaries may be due to low precision of the accurate values of their components
(this is the case also for the 70 Tau binary system).
On the other hand, hydrodynamical simulations of convection (Ludwig et al. 1999; Trampedach et al. 1999)
also give that $\alpha$ is a decreasing function of effective temperature (or stellar mass). This contradiction
needs to be explained.
}

The determination of { the} chemical composition of a star is another difficult matter.
Although it is assumed that the members of an open cluster
have the same age and chemical composition, numerous abundance determinations of various elements in
the Hyades stars do not allow the assigning of a certain value for its heavy element abundance $Z$ (see section 2).
{ Therefore, in the present study, the hydrogen abundance ($X$) and $Z$ are considered as unknowns.}

{ For the calibration of a well known binary with late-type components, the number of constraints on the
models of its components is four:
for luminosity ($L$) (or absolute magnitudes) and the radius ($R$) (or colour) of each component, we can
write down two equations:}
\begin{equation}
L_{\rm obs}=L_{\rm o}+\frac{\Delta L}{\Delta X}\delta X +\frac{\Delta L}{\Delta Z}\delta Z+
\frac{\Delta L}{\Delta \alpha}\delta \alpha+\frac{\Delta L_{}}{\Delta t}\delta t,
\end{equation}
\begin{equation}
R_{\rm obs}=R_{\rm o}+\frac{\Delta R_{}}{\Delta X}\delta X +\frac{\Delta R_{}}{\Delta Z}\delta Z+
\frac{\Delta R_{}}{\Delta \alpha}\delta \alpha+\frac{\Delta R_{}}{\Delta t}\delta t.
\end{equation}
where $L_{\rm o}$ and $R_{\rm o}$ are values { from} the reference model with fixed values of
$X$, $Z$, age ($t$) and $\alpha $ (solar values, for example, except for $t$).
{
The number of unknowns for models of the component stars is five:
 $X$, $Z$, $t$, $\alpha_{\rm A}$ and $\alpha_{\rm B}$.
Thus, if we have no extra constraint on the binary system,
 there is in principle no unique solution for it.} On the other hand,
for a visual binary in which the component stars are not well known,
we can just write two equations similar to equations (1) and (2) for total
$V$ and $B-V$ of the system.

The remainder of the present paper is organized as follows. In
Sections 2 and 3,  the observed chemical composition of the cluster
and the properties of the binary systems studied are summarized,
respectively. The model properties of the binaries and their
components are { presented in Section 4 and concluding remarks are
given in Section 5.}

\section{Chemical composition of Hyades}

There are many papers devoted to the determination of abundances of heavy elements in Hyades stars of different
classes  (from A- to K-type stars).
{ For the abundance of iron, }
Boesgaard et al. (2002) and
Hui-Bon-Hoa \& Alecian (1998) find enhancement relative to the solar abundance (0.16 dex for G-type and
0.13 for A-type stars in Hyades, respectively). {  Recently, Paulson et al. (2003) derived an abundance with a very small
formal error [(0.13$\pm $0.01 dex; see also Yong et al. 2004).]} In contrast to these findings,
Varenne and Monier (1999) find the mean
value of iron abundances from the spectra of 29 F-type stars as -0.05 dex. Thus, the abundances found from the stellar
spectra by different research groups are in general not in agreement with each other. Three dimensional calculations for the
stellar atmosphere (see, for example, Asplund et al. 2004) may solve such problems.

{
The customary consideration of iron abundance as a good tracer of total $Z$ of a star
is, however, highly debatable. Iron is not among the most abundant heavy elements, and there is no one-to-one
relation between the abundances of iron and the most abundant heavy elements (for example oxygen).
This fact can be seen, for example, in Fig. 10 of
Bensby et al. (2004): for the stars in galactic disk that have an $\left[O/H\right]$ value of about zero,
$\left[Fe/O\right]$ abundance varies between -0.4 and +0.1. Thus, it is not reasonable to take  $\left[Fe/H\right]$ $=$ $\left[O/H\right]$.
}

{
Indeed, abundances  of oxygen and nitrogen are determined by many spectroscopists.}
Takeda et al. (1998) { find} the oxygen abundance as 0.10 dex for F-stars in Hyades, while King \& Hiltgen (1996) determine
it as 0.15 dex from the spectra of the two dwarfs. Takeda et al. (1998) also give the abundance of { nitrogen} as 0.30 dex.

It could be claimed that the scattering of this amount for a given element is a result of the diffusion process,
whose rate varies from star to star. However, { contradictory} results are found for
the oxygen abundance of Hyades member HD 27561: while Garcia Lopez et al. (1993) find -0.14 dex, Clegg at al. (1981)
give 0.15 dex  as the oxygen abundance of this star.
This leads us to conclude that the difference between the abundances determined by different studies is
the result of different techniques. { Consequently,} it is a very difficult job to estimate the value of
heavy element abundance
of the Hyades. In the present paper we therefore consider Z as an unknown parameter.

\section{Observed properties of the binaries of Hyades}
\begin{table*}
\caption{
{
          The individual masses of the component stars and total V and B-V of the five binary systems.
          The slopes of the main-sequence ($S_{MBV}$) near V818 Tau and 70 Tau are computed from the data in de
          Bruijne et al. (2001) by a least-square method.  The values of the slopes computed from the binary data
          are given in  parentheses.
}
}
\label{ta5}
     $$
\begin{array}{p{0.15\linewidth}ccccccccl}
\hline
\hline
            \noalign{\smallskip}

System      & M_{\rm A}/M_\odot& M_{\rm B}/M_\odot& R_{\rm A}/R_\odot& R_{\rm B}/R_\odot&  V & B-V & S_{MBV} & \log {\frac{L_A}{L_B}}& Ref.\\
            \noalign{\smallskip}
            \hline
            \noalign{\smallskip}

V818 Tau    &1.072\pm 0.010  &0.769\pm 0.010  &{ 0.905\pm 0.029}&{ 0.773\pm 0.015}& 8.28 & 0.73 & 4.6 (4.8) & 0.71 & 1,2,3,8\\
70 Tau      &1.363\pm 0.073  &1.253\pm 0.075  & ...          & ....         & 6.46 & 0.49 & 6.6 (5.2) & .... & 4,8    \\
51 Tau      &1.80\pm 0.13    &1.46\pm 0.18    & ...          & ....         & 5.65 & 0.28 & ... & .... & 5,8         \\
$\theta^2$ Tau &2.42\pm 0.30 &2.11\pm 0.17    & ...          & ....         & 3.40 & 0.18 & ... & .... & 6,8        \\
HD 27149    &1.096\pm0.002   &1.010\pm0.002   & ...          & ....         & 7.53 & 0.68 & ... & .... & 7          \\
            \noalign{\smallskip}
            \hline
\end{array}
     $$
1) Peterson and Solensky (1988), 2) Yoss et al. (981), 3) Schiller and Milone (1987),
4) Torres et al. (1997b), 5)Torres et al. (1997c), 6) Torres et al. (1997a), 7) Tomkin (2003), 8) Lebreton et al. (2001)
\end{table*}

   \begin{figure*}
\begin{center}{\epsfbox{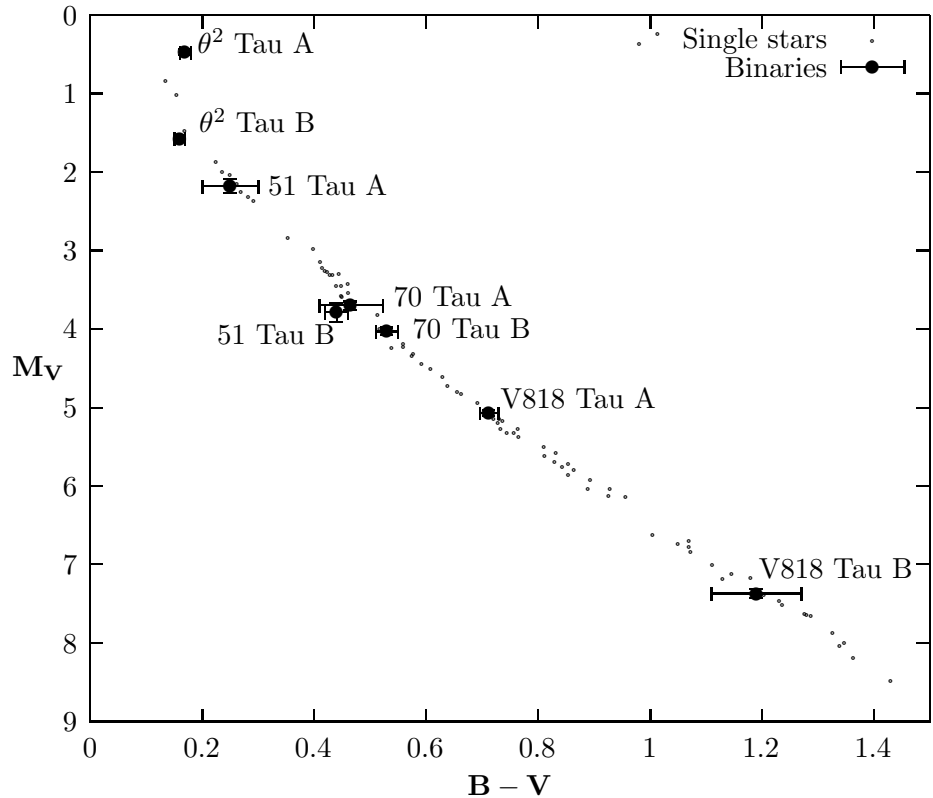}}
\end{center}
      \caption{Colour-magnitude diagram for the single stars (dots) and the components of four binaries in Hyades.}
   \end{figure*}

{
Several studies have been devoted to determining the fundamental properties of the components of V818 Tau
(McClure 1982; Schiller and Milone 1987; Peterson and Solensky 1988).
Peterson and Solensky (1988) found the masses of the primary and the secondary components
of the V818 Tau as 1.072 $\pm 0.010$ M$_{\odot}$ and 0.769 $\pm 0.005$ M$_{\odot}$, respectively.
Recently,  Torres and Ribas (2002) also found the masses of the components: $M_{\rm A}=1.0591 \pm 0.0062$ M$_\odot$ and
$M_{\rm B}=0.7605 \pm 0.0062$ M$_\odot$.
Although the individual masses found by
these two studies are very close to each other, $V$ found from the models with the former masses is in
better agreement with the observed $V$ than that of the latter. Therefore,  in our model computations
for this system, we use masses found by Peterson and Solensky (1988).
Owing to the activity feature of the system pertaining to the late-type stars, the measured values of $V$ and  $B-V$
are dispersed. Therefore, we compare theoretical results with their minimum values, $V=8.28$ and $B-V=0.73$,
observed by Yoss et al. (1981). { The radii of the components are found by Peterson and Solensky (1988) as
$R_{\rm A}=0.905\pm 0.029 R_\odot$ and $R_{\rm B}=0.773\pm 0.015 R_\odot$}. Torres and Ribas (2002) also find
very similar values to these.

70 Tau
is a close visual binary in the Hyades cluster. Torres et al.  (1997b)
determined the masses
of the primary and secondary components as  1.363 $\pm 0.073$ M$_{\odot}$ and 1.253 $\pm 0.075$ M$_{\odot}$, respectively.
Torres et al. (1997c) carried out a similar study for also 51 Tau, and obtained the masses of its
components as $M_{\rm A}=1.80$ $\pm 0.13$ M${_\odot}$ and $M_{\rm B}= 1.46$ $\pm 0.18$  M${_\odot}$.
They confirmed also thational51 Tau A is a fast rotator:
$v_{\rm A} \sin i = 97-125$ km s$^{-1}$.


$\theta^{2}$ Tau
is another spectroscopic binary systemi. Its primary component
is one of the brightest stars in Hyades and
is a $\delta$-Sct type variable star. Furthermore, the binary system consists of an evolved
and a main-sequence star. This binary system is therefore very important for testing the
different evolutionary stages. Torres et al. (1997a) found
the masses of the components to be $M_{\rm A}=2.42$ $\pm 0.30$ M${_\odot}$ and
$M_{\rm B}= 2.11$ $\pm 0.17$ M${_\odot}$.
They also found the rotational velocities of
the components. According to their results, both of the components of
$\theta^{2}$ Tau are rapid rotators: $v_{\rm A} \sin i = 80$ km s$^{-1}$ and
$v_{\rm B} \sin i = 90-170$ km s$^{-1}$.

HD 27149
is also a spectroscopic binary in the Hyades.
The minimum masses for its components have been found by Tomkin (2003) as
$M_A= 1.096\pm0.002$ M$_\odot$  and $M_B= 1.010\pm0.002$ M$_\odot$.
Tomkin confirmed that these minimum masses are unexpectedly large for the spectral type of the stars, thus
suggesting the possibility of eclipses.
By comparing appropriate models with the observations we
find the masses of the components and then the inclination angle of the system (see  section 4.4).

In Fig. 1, absolute magnitudes of the components of V818 Tau and the other binaries,
as given by Lebreton et al. (2001),
are plotted with respect to their colours. For comparison, the single stars with accurate data, given by
de Bruijne et al. (2001), are also plotted in this figure (dots). The basic data of the systems, needed for
our calibration process, are listed in Table 1.

We also use the slopes of the Hyades MS as constraints in this process.
The middle and the lower parts of the MS have different
slopes. By applying a least-square method to the data given by de Bruijne et al. (2001),
we obtain the slope of the lower part, which contains the components of V818 Tau,
\begin{equation}
S_{MS1}=\frac{\Delta M_V}{\Delta (B-V)}=4.6.
\end{equation}
Similarly, for the upper part, which contains the components of 70 Tau, we find
\begin{equation}
S_{MS2}=\frac{\Delta M_V}{\Delta (B-V)}=6.6.
\end{equation}
When we use all the available data in WEBDA {\footnote{http://www.univie.ac.at/webda}} data base, we obtain very similar results:
$S_{MS1}=4.4$ and $S_{MS2}=6.4$. However, from their absolute magnitudes and
colours, we obtain the derivative for V818 Tau (Schiller and Milone 1987) as
$S_{MS1}=4.8$ and for 70 Tau (Torres et al. 1997b) as $S_{MS2}=5.2$.
The derivative for components of 70 Tau
is significantly less than the values obtained from the data of other similar stars in
the cluster. This may arise from the fact that it is not easy to distribute
correctly the total $V$ and $B-V$ of a visual binary among its components.


}
\section{Modelling the components of the binaries of Hyades}
The characteristics of our code were already
described in Y{\i}ld{\i}z (2000;2003) (see also references therein),
and therefore we shall not provide full details here.
Our equation of state uses the approach of Mihalas et al.~(1990) in the
computation of the partition functions. The radiative
opacity is derived from Iglesias et al.~(1992), and is completed
by the low temperature tables of Alexander \& Ferguson (1994).
For the nuclear
reactions rates we use the analytic expressions given
by Caughlan \& Fowler (1988), and we employ
the standard mixing-length theory for convection (B\"{o}hm-Vitense 1958).

{
For comparison of the theoretical and observational values of the binary systems,
we compute the theoretical  $V$ and $B-V$ of any system.
We first construct models for its components with given masses and then  find $M_V$ and $B-V$
using tables for model atmospheres (Bessel  et al. 1998). Using the parallax of the binary
and $M_V$ and $B-V$ of models for the individual stars, we find the combined total $V$ and $B-V$
of the system.
}

\subsection{Solutions from V818 Tau and 70 Tau}

{
For these binary systems, we have seven equations (V and B-V of the systems 70 Tau and V818 Tau,
$L_A/L_B$ in V818 Tau, and the slopes in the middle and lower parts of the MS
in the Hyades cluster) similar to equations (1) and (2)  with seven unknowns (X, Z, t and 4 $\alpha $ for 4 stars).
For the solution of these seven equations,
we need the derivatives of seven quantities with respect to independent variables (unknowns).
These derivatives are computed
by using the reference models with
solar composition (X=0.705, Z=0.02) and $\alpha=2.0$ for all the
components at $t=1.0\times 10^9$ yr. 
The solution we find for this case is the following (Set A) \\
$X=0.679,~Z=0.0319,~\alpha_{\rm V818A}=1.89,~\alpha_{ \rm V818B}=1.01,$\\
$\alpha_{\rm 70 Tau A}=2.43,~\alpha_{\rm 70 Tau B}=2.33,~t=590$ Myr.\\
\begin{table*}
\caption{The observable quantities of V818 Tau and 70 Tau and their components computed from the models with
Set A.}
\label{ta2}
\begin{tabular}
{lllllllllll}
\hline \hline

            & L    & R   & T$_{\rm eff}$& M$_{V}$ & B-V  &U-B    & V    & B-V   &S$_{MBV}$ & $\log {\frac{L_A}{L_B}}$ \\
\hline
           &0.849 &0.950 & 5692         &5.007   &0.701 & 0.203 &8.263   &0.744    &4.603 & 0.692  \\
           &0.173 &0.741 & 4327         &7.362   &1.213 & 1.176 &        &         &      &          \\
\hline
           &2.766 &1.297 & 6543         &3.643   &0.457 & 0.020 &6.459   &0.483    &6.953 &          \\
           &1.834 &1.145 & 6285         &4.107   &0.524 & 0.010 &        &         &      &           \\
\hline
\end{tabular}
\end{table*}
The observable properties of these binary systems themselves and their components
obtained from the models with Set A { are
listed in Table \ref{ta2}. The theoretical visual magnitudes and colours } are
very close to the observed values for both systems.
}

Is the result we found from the solution of seven equations a unique one? Unfortunately, the answer is not simply yes.
One of the main reasons for this is that the numerical
{
derivatives 
are not constant}
but depend on the intervals of the variables (or on the reference model).
{
To confirm how the derivatives depend on the intervals,
we reevaluate the numerical derivatives from the models with  \\
$X=0.69,~~Z=0.032,~~~~\alpha_{\rm V818A}=2.11,~~~~\alpha_{\rm V818B}=1.3,$\\
$\alpha_{\rm 70 Tau  A}=2.66,~~~~~\alpha_{\rm 70 Tau  B}=2.45,~~t=1100$ Myr.\\
}
\begin{table*}
\caption{The observable quantities of V818 Tau and { 70 Tau} and their components computed from the models with
Set B.}
\label{ta4}
\begin{tabular}
{lllllllllll}
\hline \hline

            & L   & R    & T$_{\rm eff}$& M$_{V}$& B-V  &U-B    & V    & B-V &S$_{MBV}$ & $\log {\frac{L_A}{L_B}}$ \\
\hline
           &0.838 &0.929 &    5736      & 5.013  &0.687 & 0.182 &8.289 &0.727&4.385 & 0.730   \\
           &0.156 &0.748 &    4200      & 7.578  &1.272 & 1.241 &      &     &      &          \\
\hline
           &2.828 &1.316 &    6531      & 3.619  &0.460 &-0.018 &6.443 &0.489&6.343 &          \\
           &1.847 &1.165 &    6240      & 4.103  &0.536 & 0.021 &      &     &      &           \\
\hline
\end{tabular}
\end{table*}
{
Using these derivatives, we resolve the seven equations simultaneously and find the following values
for the seven unknowns (Set B):\\
$X=0.695,~Z=0.0298,~\alpha_{\rm V818A}=2.17,~\alpha_{\rm V818B}=0.94,$\\
$\alpha_{\rm 70 Tau A}=2.52,~\alpha_{\rm 70 Tau  B}=2.25$,~$t=1030$ Myr.\\
}
The values of the observables of the binaries found from the models with this set,
given in { Table \ref{ta4},} are in perfect agreement with the observations.
{ However we do not try to fit directly model radii to the observed radii of the components of V818 Tau.
The difference between them is very small, about 3 percent.
Indeed it is possible to remove this difference, and then the agreement between the
theoretical and observational values of B-V and  S$_{MBV}$ (the corresponding slope of the MS) of the system will disappear.}

The very striking common feature of the solutions with Set A and Set B is that
$\alpha_{\rm V818A}\sim 2 \alpha_{\rm V818B}$.
Although, both sets (Set A and B) give very similar results, the values for the age of the cluster
in both sets are very different from each other. Therefore, we shall consider
{
a star which evolves faster than these stars to fix the age; this star is $\theta^2$ Tau A.
}



\subsection{Properties of $\theta^2$ Tau and age of the cluster}
   \begin{figure*}
\begin{center}{\epsfbox{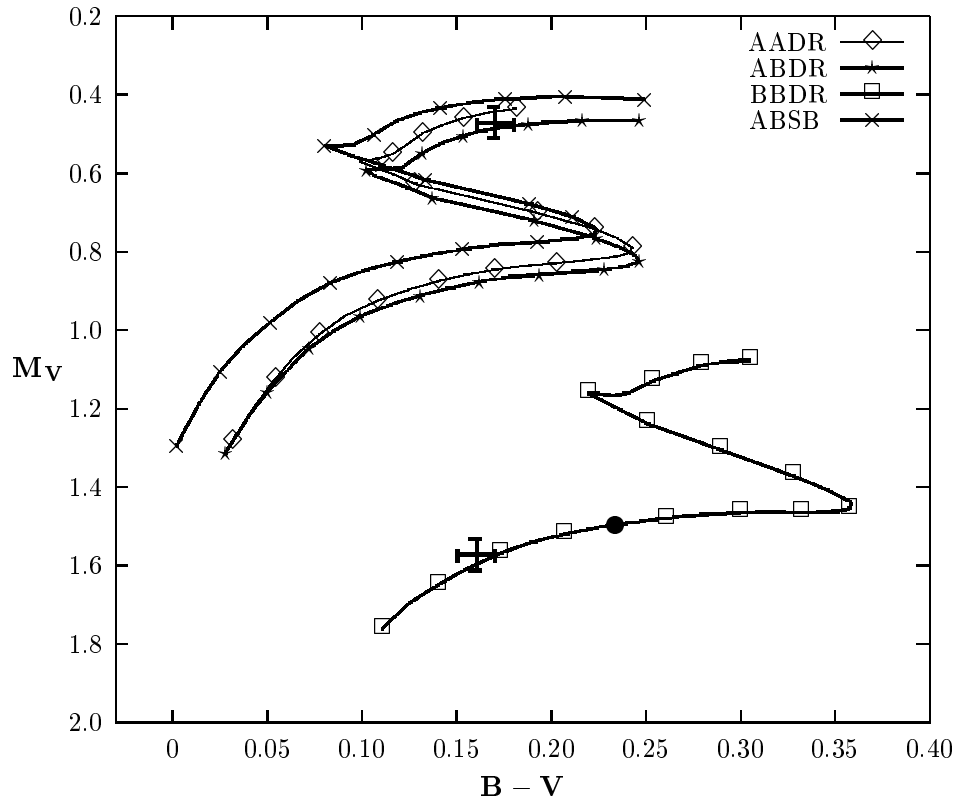}}
\end{center}
      \caption{Colour-magnitude diagram for $\theta^2$ Tau A and B. The first letter in the legend indicates
               component (A or B) and the second letter shows the chemical composition (Set A or B) of the models.
               { The }third and fourth letters specify the type of rotation (solid-body or differential rotation)
               }
              {\label{fa2}}
   \end{figure*}
{ We shall first see which one of the models of $\theta^2$ Tau A
with Set A is in better agreement with the observations. }
As mentioned above, $\theta^2$ Tau A rotates rapidly, and therefore this effect should be taken
into account. If the rotation is included in the model computations, the problem arises of angular
momentum distribution inside early-type stars.
We should first specify how inner regions of this star rotate.
Because this is a very complicated matter and is indeed one of the essential problems in stellar astrophysics,
we study two typical cases:
i) solid rotation, ii) the rotation profile as determined by contraction. The models with the latter case are
differentially rotating (DR) models: the central regions are rotating much faster
than the outer regions { (Y{\i}ld{\i}z 2003; 2004). }

In Fig. \ref{fa2}, the DR models of $\theta^2$ Tau A with Set A (Model AADR; diamond) and Set B (ABDR; star), and the
model rotating
like a solid body (ABSB; $\times$) are plotted in the HR diagram. Their $B-V$ values are in good agreement with the observed
one (0.17) (de Bruijne et al. 2001) at $t=675$ Myr, $699$ Myr and $661$ Myr, respectively.
Their equatorial velocities at their corresponding ages  are $v_{eq}\sim 100$ km s$^{-1}$.
It seems that
the sub-giant phase of $\theta^2$ Tau A is compatible with the observed position and the DR models with
both sets of chemical composition  are in much better agreement with the observed values
than the models rotating like a solid body.
{
Even though the
}
solid-body rotation can not be totally ruled out by such an analysis,
we shall consider only the DR models hereafter.
What is important here in our analysis is that the ages of models with the two rotation types
are very similar to each other.

The evolutionary track of the DR model of $\theta^2$ Tau B with Set B is also plotted in Fig. \ref{fa2}.  Although
{ its evolutionary}
track passes through the observed position of $\theta^2$ Tau B in the HR diagram,
{ the time of agreement }
($t=444$ Myr) is not the same as that of $\theta^2$ Tau A ($t=699$ Myr).  The position
of $\theta^2$ Tau B at the latter time is marked by filled circle, and
is far from the observed position of $\theta^2$ Tau B in the diagram.
{
An internal rotation, more complicated than the
rotation as determined by the contraction,  may cause this discrepancy.
}
\subsection{Solutions derived from models of $\theta^2$ Tau A and 70 Tau A \& B}
As shown in the previous sections, two models having the same mass but different chemical compositions
may have the same (or very near) location in the HR diagram.
Because of this  degeneracy,
we calibrate models of $\theta^2$ Tau A and the components of 70 Tau to the observations for the fixed values of Z.
For $Z=0.028$, we obtain (Set 28)\\
$X=0.699,~\alpha_{\rm 70 Tau A}=2.31,~\alpha_{\rm 70 Tau  B}=2.21$,~$t=705$ Myr.\\
We made similar computations also for $Z=0.033$ and obtain the following results (Set 33)\\
$X=0.676,~\alpha_{\rm 70 Tau A}=2.29,~\alpha_{\rm 70 Tau  B}=2.20,~t=676$ Myr.\\
For $Z=0.024$, using the values of $~\alpha$ in Set 28, we find
$X=0.718$ and $t=721$ Myr (Set 24).

The differences between the models of any star with different
sets are negligibly
small. Therefore, these sets are equivalent { to} each other { (see Table 4)}.

We also build models ({ with} Set 28) with the microscopic diffusion process for { 70 Tau A}
to test its influence on the
observable properties of such stars.
While the difference between the absolute magnitudes of the models with and without diffusion is
{
$\Delta M_V= -0.0022$,
the difference between the colours is
$\Delta (B-V) =  0.0031$.
These differences are small enough, in comparison to the uncertainty in the observed magnitude and colour of the system,
that the diffusion process can be ignored.}

\begin{figure*}
\begin{center}{\epsfbox{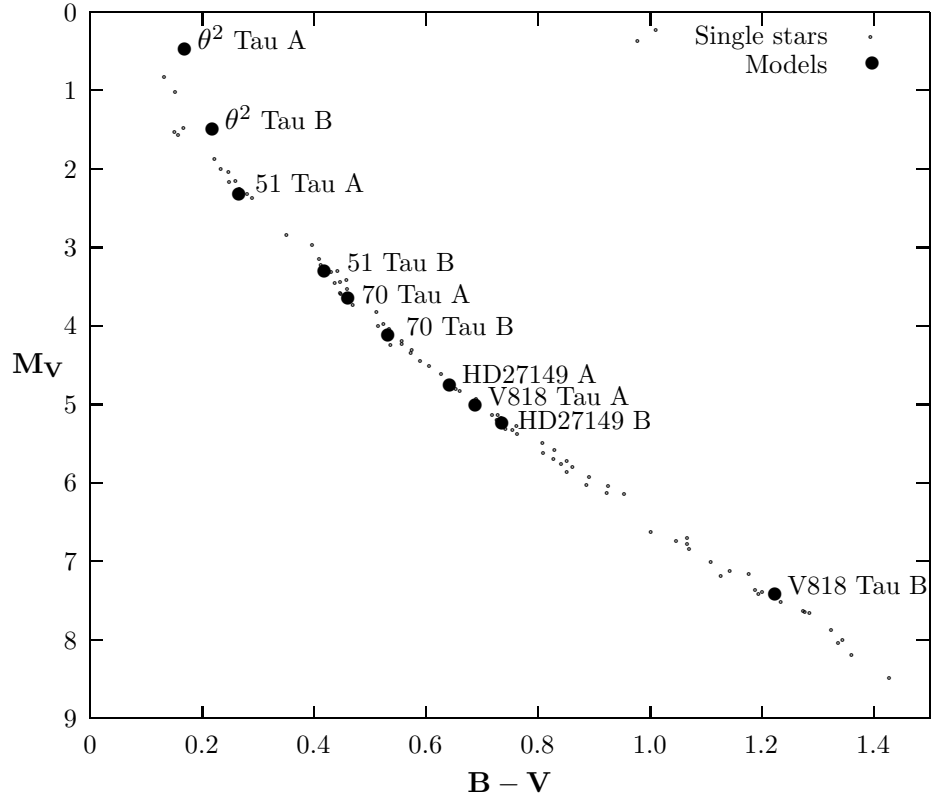}}
\end{center}
      \caption{Colour-magnitude diagram for the single stars (dots) and the models of the components of
              the five binaries { (filled circles)}.}
              {\label{f1.3}}
   \end{figure*}

For these sets, we also find that the mixing-length parameter for the components of V818 Tau are as
$\alpha_{\rm V818A}=2.04$ and $\alpha_{\rm V818B}=0.99$. In Fig. \ref{f1.3}, the models for the components
of $\theta^2$ Tau, 70 Tau and V818 Tau
with Set 24 (filled circles) are plotted in the HR diagram among the single stars with very precise parallaxes (dots).
Except for $\theta^2$ Tau B (see above), all the models of the components are qualitatively in good agreement with
the general trend identified by the single stars with very accurate observational data.

{ The} models for the components of 51 Tau with Set 24 are also placed in Fig. \ref{f1.3}.
Their positions are good enough in comparison with the positions of the single stars
(de Bruijne et al. 2001). The model of 51 Tau A is DR model with equatorial velocity of 130 km s$^{-1}$, and
the mixing-length parameter for 51 Tau B is 2.35.

\subsection{Properties of HD 27149}
Because the angle between the line of sight and the orbital plane of HD 27149
is observationally not known, we find the individual masses of the components
by fitting $V$ of the models for a given chemical composition and time to
$V_{\rm obs}$ of the system. For Set 24 and Set 28, we find that $M_{\rm A}=1.118$
and $M_{\rm B}=1.030$. Then, from the calibration of models of these masses
with both sets, the mixing-length parameters are derived as { $\alpha_{\rm A}=2.10$ and $\alpha_{\rm B}=1.95$.}
These different values of the mixing-length parameter for the different stellar masses prove once more the mass dependence
of $\alpha$.

From these values of individual masses, we can obtain the value of the inclination angle by using the observed
value of $M_{\rm A}\sin^3 i$ (Tomkin, 2003):
\begin{equation}
M_{\rm A}\sin^3 i=1.118 \sin^3 i=1.096.
\end{equation}
Equation ({5}) gives the inclination of the system as $i=83^{\circ}.4$.
The minimum value of $i$ for eclipsing to occur is, however,  $i_c=88^{\circ}.9$.
{
Thus,
HD 27149 is not an eclipsing binary (see Tomkin, 2003).}

\subsection{The stellar mass dependence of the mixing-length parameter}
\begin{table*}
\caption{The mixing-length parameters of the calibrated models of the components of binaries V818 Tau,
         70 Tau and HD 27149 with Set 24. In order to derive a fitting formula, models
         are also computed for the masses 0.80, 0.85,0.90 and 0.95 M$_\odot$
          (see Fig. 4).
         The uncertainties in $\alpha$ of each model are computed assuming an uncertainty of $\Delta (B-V) =  0.005$
         (see the text). { For comparison, radii and effective temperatures of the models with Set 24 and Set 28
          are also given. The fractional differences between the model radii of components of V818 Tau and 70 Tau with these          sets are
          listed in the last column.
         }
         }
\label{ta5}
\begin{tabular}
{llcccccc}
\hline
\hline

Star      & $M/M_\odot$&$\alpha$  &$T_{\rm eff}$(Set 24)&$R/R_\odot$(Set 24)&$T_{\rm eff}$(Set 28)&$R/R_\odot$(Set 28)&$\delta R/R$\\
\hline
V818 Tau B   &0.769   &0.99 $\pm$ 0.03 &   4305    & 0.739        & 4295         &    0.741    & 0.003 \\
V818 Tau A   &1.072   &2.04 $\pm$ 0.09 &   5718    & 0.940        & 5720         &    0.937    &-0.003   \\
70 Tau B     &1.253   &2.21 $\pm$ 0.09 &   6252    & 1.154        & 6248         &    1.162    & 0.007\\
70 Tau A     &1.363   &2.31 $\pm$ 0.18 &   6527    & 1.306        & 6514         &    1.318    & 0.009\\
HD 27149   B &1.030   &1.95 $\pm$ 0.07 &   5583    & 0.896        & ...         &    ...    &  ...   \\
HD 27149   A &1.118   &2.10 $\pm$ 0.08 &   5876    & 0.985        & ...         &    ...    &  ...     \\
--           &0.800   &1.20 $\pm$ 0.05 &   4500    & 0.749        & ...         &    ...    &  ...     \\
--           &0.850   &1.50 $\pm$ 0.06 &   4764    & 0.768        & ...         &    ...    &  ...     \\
--           &0.900   &1.70 $\pm$ 0.06 &   5086    & 0.799        & ...         &    ...    &  ...     \\
--           &0.950   &1.80 $\pm$ 0.07 &   5329    & 0.836        & ...         &    ...    &  ...     \\
Sun          &1.000   &1.88      \\
\hline
\end{tabular}
\end{table*}

   \begin{figure*}
\begin{center}{\epsfbox{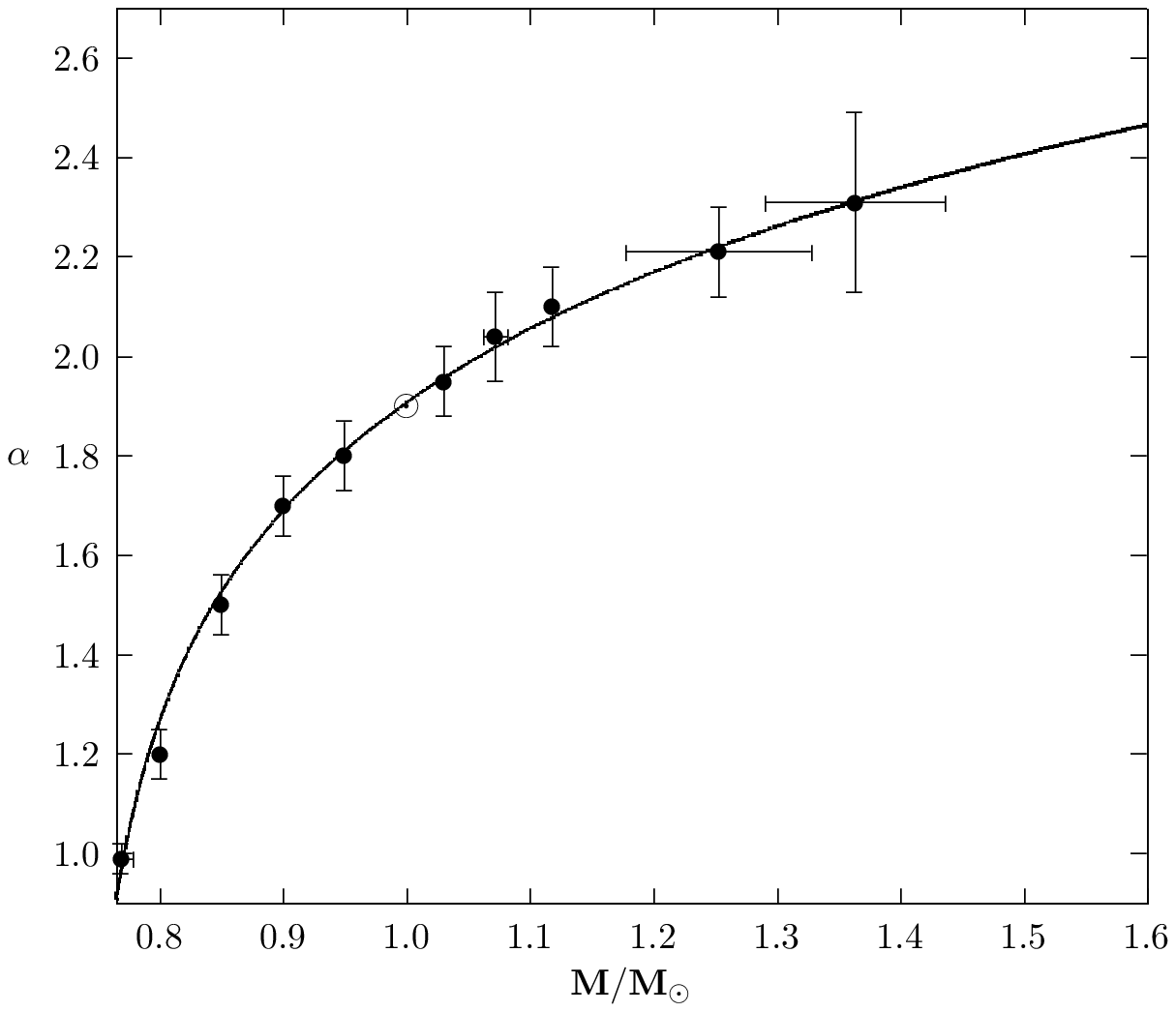}}
\end{center}
      \caption{The mixing-length parameter for the models of the components of { the three} binaries
               (V818 Tau, 70 Tau and HD 27149) as
               a function of mass.
               In order to fill the gap, models are computed also for the masses 0.80, 0.85,0.90 and
               0.95 M$_\odot$ by taking care to keep the model in the sequence of stars with data of high
               quality in the CM diagram. { For the method of computation for the uncertainty in $\alpha$,
               see the text.}
                }
              {\label{f1.4}}
   \end{figure*}
The values of the mixing-length parameter for the components of the binaries
derived from the calibration are listed in Table \ref{ta5}. They are
plotted in Fig. 4 as a function of the stellar masses.
The uncertainties
in $\alpha$ of each model are computed assuming an uncertainty of $\Delta (B-V) =  0.01$ { (uncertainty in B-V of the binary systems; see Lebreton et al. 2001)} for each star.
Then,
\begin{equation}
\delta \log \alpha = \frac{\Delta (B-V)}{ {\partial (B-V)}/{\partial \log \alpha }}
\end{equation}
where the partial derivatives are computed from the models. {
In Table 4, radii and effective temperatures of the models with Set 24 and Set 28 are also listed.
In the last column, the fractional difference between the models with these sets are given
for the components of V818 Tau and 70 Tau. Because these differences are very small, we deduce that
the mass dependence of $\alpha$ is independent of the (solution) sets.
}

The curve {  in Fig. 4} is the fitting formula,
\begin{equation}
\alpha = 9.19 (M/M_\odot-0.74)^{0.053}-6.65
\label{eq7}
\end{equation}
 derived using the data given in { Table \ref{ta5}}.
It is valid for the stellar masses greater than $0.77M_\odot$.
The curve surprisingly { also} covers $\alpha$ of the Sun ($\odot$ in Fig. 4).
{ Moreover, $\alpha$s of the components of $\alpha$ Centauri from equation (7) explain the
stellar parameters of these stars well (Y{\i}ld{\i}z, in preparation).
}

   \begin{figure*}
\begin{center}{\epsfbox{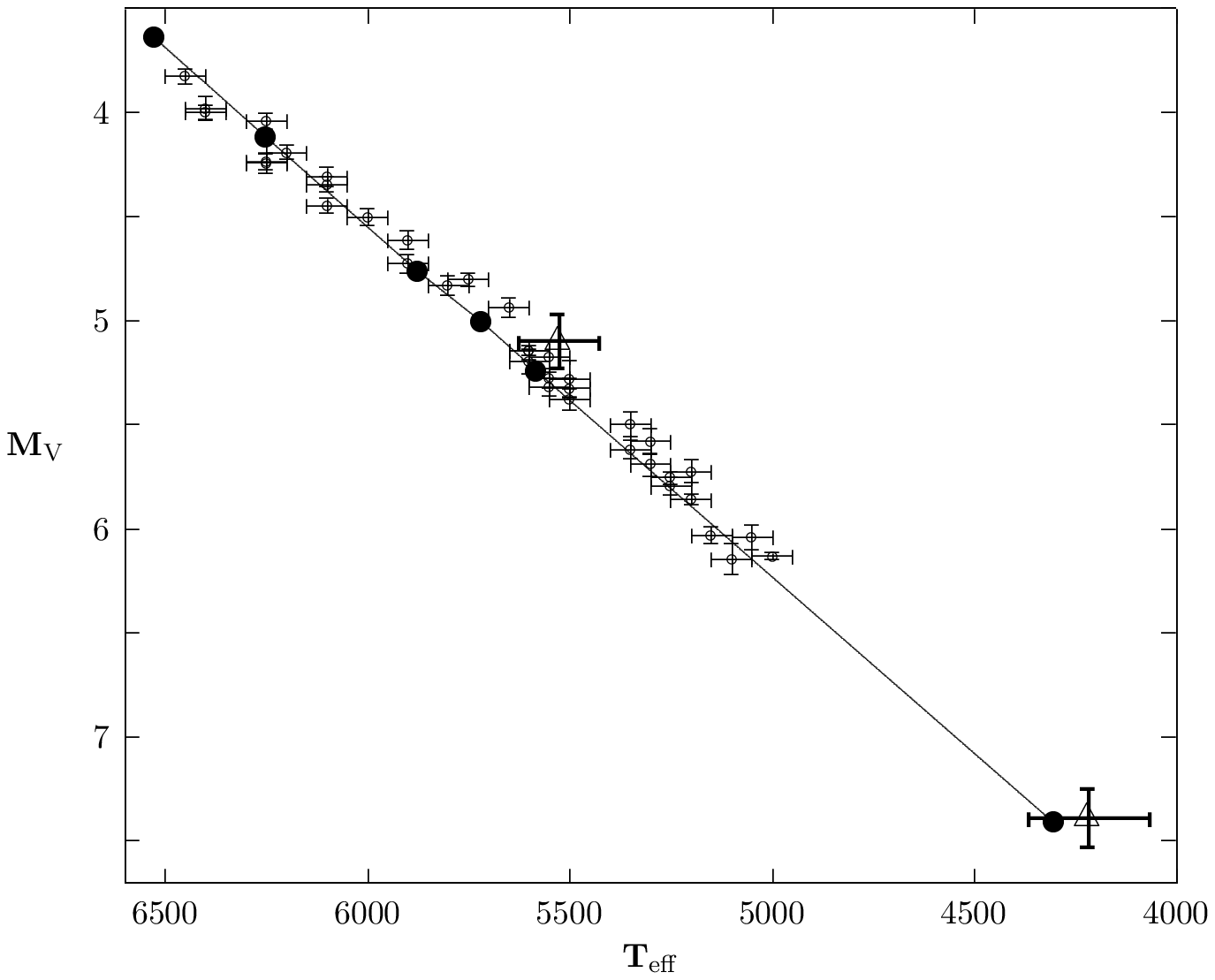}}
\end{center}
      \caption{  The models of the components of { the three} binaries
               (V818 Tau, 70 Tau and HD 27149) with Set  24 (filled circles) are plotted in
               $M_{\rm V}$ vs. $T_{\rm eff}$ diagram. The error bars with circles represent the observations:
               while the effective temperatures are spectroscopicallly derived by Paulson et al. (2003),
               the absolute magnitudes are from de Bruijne et al. (2001). The error bars with triangles
               are for the components of V818 Tau based on the photometric data (Torres and Ribas, 2002)
                }
              {\label{f1.4aa}}
   \end{figure*}
{
One might also want to ascertain whether the effective temperatures of the models with variable $\alpha$
are consistent with spectroscopic measurements. For this task, similar to fig. 2 in  Pinsonneault et al. (2004),
we plot $M_{\rm V}$ of the models (Set  24) of the late type components of the three binary systems against  the
$T_{\rm eff}$ (filled circles) in Fig. 5. The observational effective temperatures and absolute magnitudes
are taken from Paulson et al. (2003) and de Bruijne et al. (2001), respectively. We include also photometric
data (triangle with thick error bars) for the components
of V818 Tau (Torres and Ribas, 2002), because of a lack of observational data for stars with a $T_{\rm eff}$
less than 5000 K. It can be  seen that the theoretical and the observational effective temperatures are in very good agreement.

}

   \begin{figure*}
\begin{center}{\epsfbox{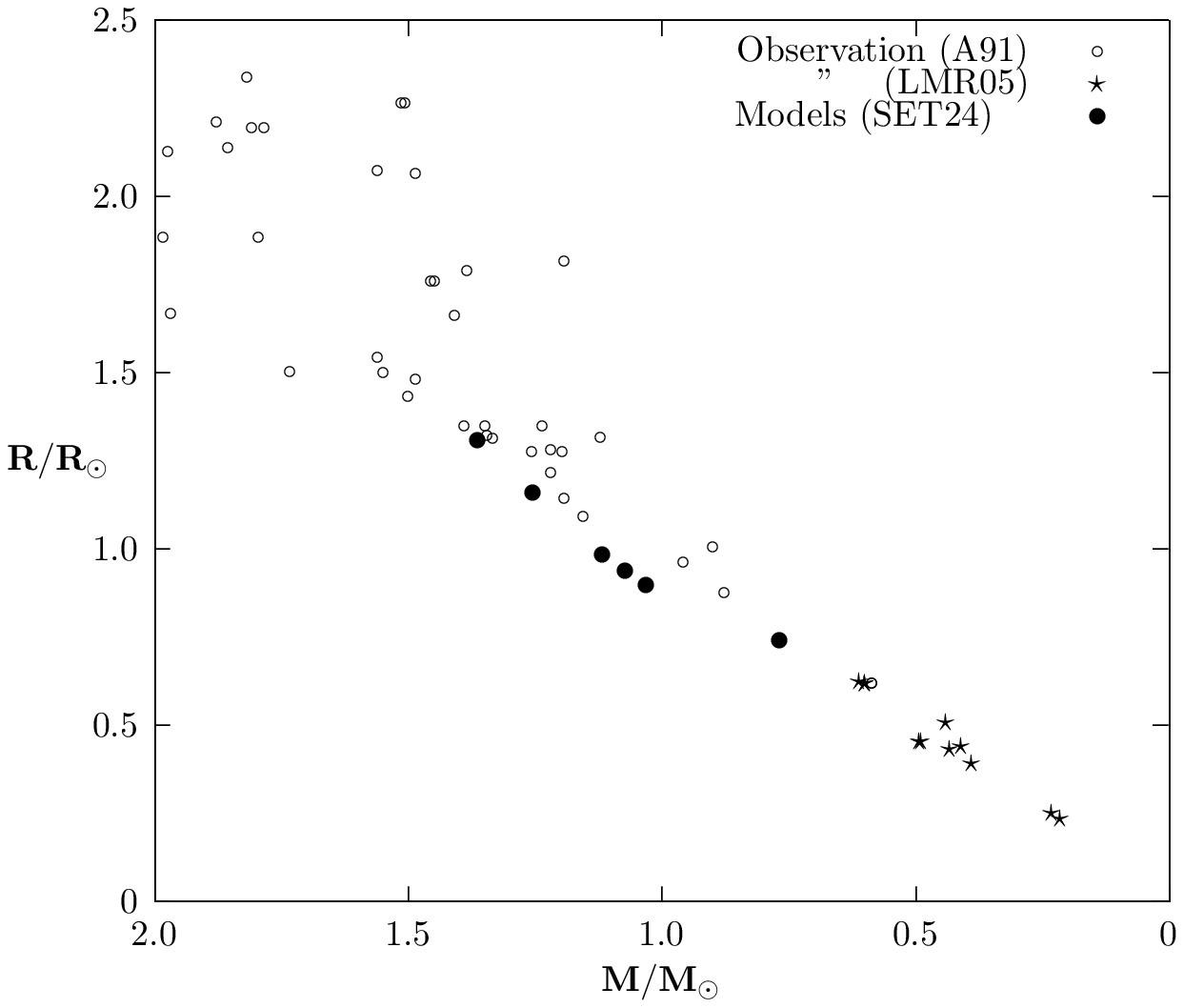}}
\end{center}
      \caption{  The models of the components of { the three} (V818 Tau, 70 Tau and HD 27149) binaries with Set  24 (filled circles) are plotted in
               a $R_{\rm }$ vs. $M_{\rm }$ diagram.
               The circles and stars represent the observed stellar radii taken from Andersen (1991) and
               Lopez-Morales and Ribas (2005), respectively.
                }
              {\label{f1.4a}}
   \end{figure*}

We also compare the model radii
with the observed radii of stars in well known eclipsing binaries. In Fig 6, the observed radii, taken from
Andersen (1991) (circles) and Lopez-Morales and Ribas (2005) (stars), are plotted against the stellar mass.
The filled circles represent the models of the components of  V818 Tau, 70 Tau and HD 27149  with Set 24.
We confirm a very good agreement between the observed and the model radii.
The models for the $unevolved$ late-type stars in the Hyades cluster are very near to zero-age-main-sequence line
and therefore takes place on the left side of the main-sequence formed by the stars in the well known eclipsing binaries.

\subsection{The { colour-colour} diagram}

{
In { Fig. 7}, $U-B$ and $B-V$ colours derived from our models with Set 24 are plotted. For comparison, the observed
colours for the Hyades stars are also plotted. The filled circles and boxes represent
the colours computed from the tables of Bessel et al. (1998) and Lejeune et al. (1998),
respectively, for solar composition. Although both of the tabels are in general in good agreement with the observation, the $U-B$ colours of models of the early type stars (51 Tau A, $\theta^2$ Tau A and B) derived from Lejeune et al. (1998) are in better agreement with
the observed $U-B$ than those derived from Bessel et al. (1998). The colours of some models are also computed from Lejeune et al. (1998) for the metallicity
given as Paulson et al. (2003), namely $\left[Fe/O\right]$=0.13 dex (triangles).  As the metallicity of a model is increased, the model
moves towards the bottom-right  part of the colour-colour diagram, as expected.
The majority of the stars are located between
the colours of the models derived for the solar composition and for the metal-rich composition, at least for stars
with $B-V < 0.6$. This result may be interpreted as meaning that the Hyades cluster is slightly more metal rich
than the Sun.  More precise results on the metallicity of the cluster depend on which solar mixture is used.
Therefore, new tables with the recently calculated solar composition (Asplund et al. 2004) for colours of stars are required.
}

   \begin{figure*}
\begin{center}{\epsfbox{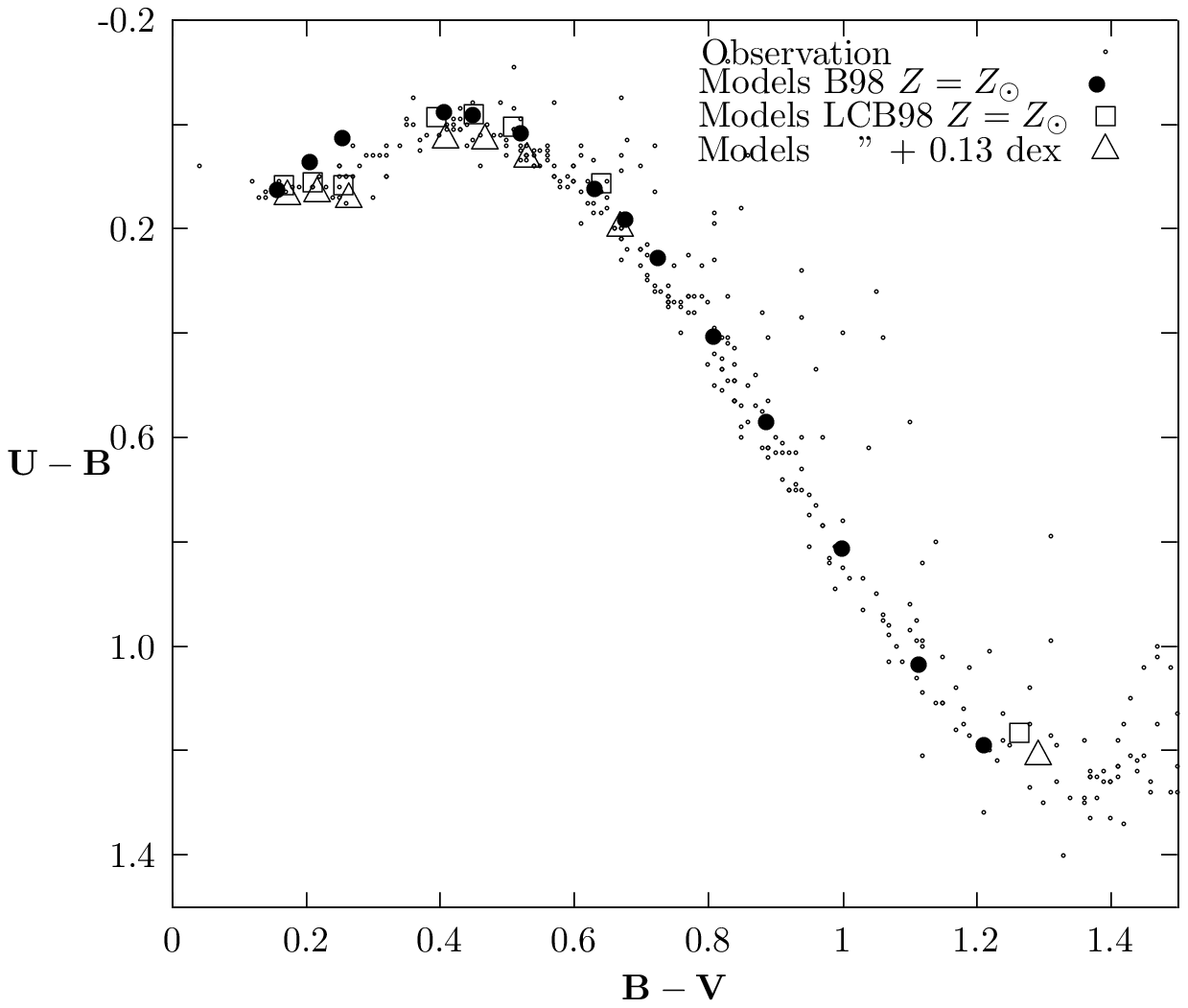}}
\end{center}
      \caption{Colour-colour diagram for the Hyades cluster. The models of the components of V818 Tau, 70 Tau and HD 27149 are
      constructed with Set 24. The filled circles represent the colours of the model computed from Bessel (1998)
      and the box is for the colours from Lejeune et al. (1998) for the solar composition. The triangle is for the
      colours from Lejeune et al.  (1998) for the metallicity [Fe/O]= 0.13 dex.
               }
              {\label{f1.6}}
   \end{figure*}

\section{Conclusion}
By constructing models for the components of the Hyades members binary systems whose masses are
observationally determined, we have reached very important conclusion concerning the detailed
physical structure as well as the fundamental properties of the cluster itself.
The most striking outcome of the present study is that we discover a smooth
       dependence of the mixing-length parameter on the stellar mass.
{ Although we calibrate the models in order to obtain the measured quantities of the corresponding binary system
rather than those of the individual component stars, the models of each star are in good agreement with the
observed properties:
while the difference between the model and data radii is about 3-4 percent for the components of V818 Tau,
the effective temperatures of the models of late-type components of V818 Tau, HD 27149 and 70 Tau
are in perfect agreement with $T_{\rm eff}$s derived from spectroscopic measurements by Paulson et al. (2003).
Because V818 Tau is an active binary, the fundamental properties of its components should be re-evaluated using
much more precise observational data than are used at present.}

The relationship we derived between $\alpha$ of the components of
the Hyades' binaries and their masses also gives the solar value of
$\alpha$. This result should, however, be examined. It is possible
that the age and chemical composition differences just happen to
counterbalance each other, and that therefore the relationship gives
the solar value coincidentally. The fact that the $\alpha$ values
for the components of $\alpha$ Centauri  from this relationship
yield models in agreement with the observed stellar parameters leads
us to adopt it as a prevalent relationship (Y{\i}ld{\i}z, in
preparation). However, time variation of this relationship should
not be ruled out.

From 2D and 3D hydrodynamical simulations of convection, Ludwig et
al. (1999) and Trampedach et al. (1999) find that $\alpha$ is a
decreasing function of effective temperature (or stellar mass). It
is noteworthy that stellar evolution codes and simulation codes give
opposite results for this relationship (see also Ludwig and Salaris
1999 for $\alpha$ values of the components in the eclipsing binary
AI Phe). { This contradiction may be a result of the variation of
$\alpha$ in the layers of convective envelopes (see e.g., Deupree
and Varner 1980) or in time. }

The second important outcome concerns structure of the early type
stars: { the} differentially rotating models for the components of
$\theta^2$ Tau and 51 Tau are in better agreement with the
observations than the non-rotating models and models with solid-body
rotation.

The fundamental properties of the Hyades cluster that we have derived are not unique but can be given in terms of
the metal abundance. If $Z=0.024$, then $X=0.718$ and its age is $t=721$ Myr; if $Z=0.033$, then $X=0.676$ and $t=676$ Myr.
It should be pointed out here, however, that the mass dependence of $\alpha$ is valid regardless of the value
assigned for $Z$.

From these fundamental properties of the cluster we derive the masses of the components in the binary
system HD 27149 by fitting the brightness of the models to the observed value: $M_{\rm A}=1.118 M_\odot$ and
$M_{\rm B}=1.03 M_\odot$. Using the observed lower value for the masses, we show that the inclination of its orbit is
about $i=83\degr.4$.
This value of $i$ is smaller than the critical value for the occurrence of eclipsing ($i_{\rm cr}=88\degr.9$) and hence
this system is not an eclipsing one.

Unless the metal abundance of the cluster is observationally
determined very precisely, we can not specify its helium abundance.
In order to be able do this, abundances of the most abundant
chemical species, in particular, oxygen, nitrogen, carbon and neon,
should be found from the spectrum of its stars. Otherwise,  we may
only give the helium abundance as a function of $Z$ as well: while
$Y=0.258$ for $Z=0.024$, $Y=0.291$ for $Z=0.033$.  With very precise
data on the colours of the stars, it is possible, however, to
specify the heavy element abundance from the colour-colour diagram.

\section*{Acknowledgments}
We thank Ay\c{s}e Lahur K{\i}rtun\c{c} and { Rachel Drummond} for their suggestions
which improved the language of the manuscript. The anonymous referee is acknowledged
for her/his useful comments. This research was supported by The Scientific and
Technological Research Council of Turkey (T\"UB\.ITAK).

\def \apj#1#2{ApJ,~{#1}, #2}
\def \aj#1#2{AJ,~{#1}, #2}
\def \astroa#1{astro-ph/~{#1}}
\def \pr#1#2{Phys.~Rev.,~{#1}, #2}
\def \prt#1#2{Phys.~Rep.,~{#1}, #2}
\def \rmp#1#2{Rev. Mod. Phys.,~{#1}, #2}
\def \pt#1#2{Phys.~Today.,~{#1}, #2}
\def \pra#1#2{Phys.~Rev.,~{ A}~{#1}, #2}
\def \asap#1#2{A\&A,~{#1}, #2}
\def \aandar#1#2{A\&AR,~{#1}, #2}
\def \apss#1#2{~Ap\&SS,~{{#1}}, #2}
\def \asaps#1#2{A\&AS,~{#1}, #2}
\def \arasap#1#2{ARA\&A,~{#1}, #2}
\def \pf#1#2{Phys.~~Fluids,~{#1}, #2}
\def \apjs#1#2{ApJS,~{#1}, #2}
\def \pasj#1#2{PASJ,~{#1}, #2}
\def \mnras#1#2{MNRAS,~{#1}, #2}
\def \ibvs#1{IBVS,~No.~{#1}}

\end{document}